\documentclass[11pt,preprint]{aastex}

\def\cc{cm$^{-3}$}

\def\be{\begin{equation}}
\def\ee{\end{equation}}
\def\sT{\sigma_{\rm T}}

\def\Rdec{R_{\rm dec}}
\def\ndec{n_{\rm dec}}

\def\Gej{\Gamma_{\rm ej}}

\def\tobs{t_{\rm obs}}
\def\texp{t_{\rm exp}}

\def\epB{\epsilon_B}
\def\epe{\epsilon_e}
\def\ep{\varepsilon}
\def\epIC{\varepsilon_{\rm IC}}
\def\tIC{t_{\rm IC}}
\def\nufluid{\nu_{\rm fluid}}
\def\epfluid{\varepsilon_{\rm fluid}}
\def\taugg{\tau_{\gamma\gamma}}

\newbox\grsign \setbox\grsign=\hbox{$>$} \newdimen\grdimen \grdimen=\ht\grsign
\newbox\simlessbox \newbox\simgreatbox \newbox\simpropbox
\setbox\simgreatbox=\hbox{\raise.5ex\hbox{$>$}\llap
     {\lower.5ex\hbox{$\sim$}}}\ht1=\grdimen\dp1=0pt
\setbox\simlessbox=\hbox{\raise.5ex\hbox{$<$}\llap
     {\lower.5ex\hbox{$\sim$}}}\ht2=\grdimen\dp2=0pt
\setbox\simpropbox=\hbox{\raise.5ex\hbox{$\propto$}\llap
     {\lower.5ex\hbox{$\sim$}}}\ht2=\grdimen\dp2=0pt
\def\simgt{\mathrel{\copy\simgreatbox}}
\def\simlt{\mathrel{\copy\simlessbox}}

\begin{document}

\title{Optical and GeV-TeV flashes from gamma-ray bursts}

 \author{Andrei M. Beloborodov\altaffilmark{1}} 
 \affil{Physics Department and Columbia Astrophysics Laboratory \\
  Columbia University, 538  West 120th, Street New York, NY 10027}

 \altaffiltext{1}{Also at Astro-Space Center of Lebedev Physical 
 Institute, Profsojuznaja 84/32, Moscow 117810, Russia}

\begin{abstract}
The synchrotron optical flash caught in GRB~990123 overlaps with the 
MeV radiation front, and the optical-emitting electrons must also produce 
GeV-TeV emission by inverse Compton scattering of MeV photons. 
The ultra-high-energy flash can be much stronger than its optical 
counterpart.
We also note that Compton cooling by MeV photons immediately terminates 
the optical emission unless the fireball Lorentz factor exceeds $10^3$.
Severe Compton losses may 
explain the non-detections of optical flashes in several long GRBs.
Such failed optical flashes should be especially efficient GeV producers 
and likely to develop $e^\pm$ cascades. This probably happened in 
GRB~941017 and its mysterious high-energy component is well explained by 
Compton upscattering of GRB photons at the fireball deceleration radius. 
The proposed mechanism of GeV emission should not work for short $\gamma$-ray
bursts that early decouple from the fireball and avoid interaction with 
the electrons in the deceleration flash. Observations by {\it Swift} and 
GLAST will provide an opportunity to test these expectations. The existing 
data for GRB~990123 already impose interesting constraints on the explosion. 
\end{abstract}

\keywords{ cosmology: miscellaneous ---
  gamma rays: bursts ---
  radiation mechanisms: nonthermal ----
  shock waves}


\section{Introduction}

Prompt optical observations of gamma-ray bursts (GRBs) are challenging 
because they require a quick pointing of an optical telescope or patroling 
the whole sky with good angular and temporal resolutions. Nevertheless,
about ten bursts have been observed in $10-100$~s time by ROTSE
instrument (Akerlof et al. 2000; Kehoe et al. 2001).
In only one of them, GRB~990123, a bright optical flash was detected, 
which reached a peak of 9th magnitude at 40-50~s after the beginning of
the GRB (Akerlof et al. 1999). The peak overlapped with the main MeV burst, 
however, was interpreted as a separate emission component because it had a 
different light curve and showed a tail $\sim 10$ times longer than the 
MeV burst. Similar tails of optical flashes were caught in a few other 
bursts at times less than $10^3$~s (Fox 2002; Fox et al. 2003).
Such flashes are expected from the reverse shock in the GRB fireball 
(M\'eszaros \& Rees 1993; Sari \& Piran 1999) or the early forward shock 
in the external medium (Beloborodov 2004, in preparation). 
It was unclear, however, why they were not detected in most of the 
bursts observed by ROTSE (Akerlof et al. 2000; Kehoe et al. 2001), 
and possible reasons have been discussed (e.g. Kobayashi 2000;
Nakar \& Piran 2004).

\section{Optical flash}

The optical flash is interpreted as synchrotron emission of 
relativistic electrons injected in a magnetic field $B$.
Emission from electrons with Lorentz factors $\gamma_e$ in the 
fluid frame peaks at frequency $\nufluid \approx 0.2(eB/m_ec)\gamma_e^2$
(assuming isotropic pitch-angle distribution). The corresponding observed 
frequency is modified by the Doppler effect and a cosmological redshift 
of the burst, $\nu\approx(1+z)^{-1}\Gamma\nufluid$, where $\Gamma$ is the 
fluid Lorentz factor and $z$ is the redshift. The characteristic $\gamma_e$ 
that gives emission at the observed frequency $\nu$ is
\be
\label{eq:tgnu}
  \gamma_e(\nu)=\left[\frac{5\,m_ec\nu(1+z)}{\Gamma eB}\right]^{1/2}.
\ee
The energy density of the magnetic field, $w_B=B^2/8\pi$, is a fraction 
of the total plasma energy in the emission region, $w$.
As is customary, we parameterize $B$ by $w_B=\epsilon_B w$.

If the flash is caused by the fireball interaction with an ambient medium,
$w$ is estimated from the jump condition 
at the forward shock: $w=4\Gamma^2 nm_pc^2$ where $n$ is the ambient 
density. The energy density is about this value everywhere between the 
forward and reverse shocks (while $\epB$ may be significantly different
on the two sides of the contact discontinuity). One then finds 
\be
\label{ge_bw}
 \gamma_e(\nu)\approx 3\times 10^2 (1+z)^{1/2}\nu_{15}^{1/2}
                      \Gamma_2^{-1}(\epB n)^{-1/4},
\ee
where $n$ is expressed in \cc.

The observed flash in GRB~990123 emits $\sim 10^{-3}$ of the GRB energy 
and then decays with time as a power-law of index $1.6-2$, which extends 
to at least $10^3$~s. This decay was interpreted as a result of adiabatic 
cooling of the injected relativistic electrons in the expanding fireball, 
which requires the synchrotron cooling to be relatively slow,
\be
\label{eq:sc}
  t_s(\gamma_e)=\frac{3m_ec}{4\sT w_B\gamma_e}>\texp=\frac{R}{c\Gamma}.
\ee
Here $t_s$ is the synchrotron cooling timescale and $\texp$ is the
expansion time at a radius $R$ (the timescale of adiabatic cooling); 
$t_s$ and $\texp$ are measured in the fluid frame. 
Using equation~(\ref{ge_bw}) this condition can be rewritten as
\be
\label{bound1}
  \epB n < 2\times 10^{-2}\,(1+z)^{-2/3}\nu_{15}^{-2/3} R_{17}^{-4/3}.
\ee

\section{Inverse Compton cooling by MeV radiation}

The peak of the optical flash in GRB~990123 arrived in the middle of 
the prompt burst. It implies that most of the flash-emitting electrons 
were exposed to the MeV photons. This is so even if the MeV
source is ``patchy'' --- the 
photons would propagate and fill the observable part of the 
fireball $R/\Gamma$ where the flash could come from.

\subsection{The slow-cooling condition}

The energy density of the $0.1-1$~MeV photons in the flash region is
given by 
\be
   w_\gamma=\frac{E_\gamma}{4\pi R^2\Delta\Gamma^2},
\ee
where $E_\gamma$ is the isotropic equivalent of the burst energy,
$\Delta=(1+z)^{-1}ct_b$ is the thickness of MeV radiation front, 
and $t_b$ is the observed GRB duration. The radiation density $w_\gamma$
is measured in the fluid frame where the photon energy is
$\epfluid=\ep/\Gamma\sim {\rm keV}$.
Electrons with $\gamma_e(\nu)\sim 10^2-10^3$ will upscatter the photons 
with Thomson cross section if $\epfluid<m_ec^2/\gamma_e(\nu)$, which is 
comparable or exceeding the main peak of GRB spectrum, and therefore a 
significant fraction of $w_\gamma$ will efficiently cool the flash-emitting 
electrons. A slow-cooling model of the flash must satisfy the condition,
\be
   \tIC(\gamma_e)=\frac{3m_ec}{4\sT w_\gamma\gamma_e} > \texp.
\ee
We note that Compton cooling by optical radiation is much weaker compared 
with the upscattering of MeV photons because the energy of optical 
radiation is relatively small: $w_O\sim 10^{-3}w_\gamma$ in GRB~990123.

Substitution of $\gamma_e(\nu)$ from equation~(2) gives
\be
   \frac{\tIC(\nu)}{\texp}=4\times 10^{-3}\,\Gamma_2^4\,\nu_{15}^{-1/2}
   (1+z)^{-1/2}(\epB n)^{1/4}E_{\gamma,54}^{-1}\Delta_{12}R_{17}.
\ee
The flash in GRB~990123 peaks on a short timescale $\tobs <100$~s and 
therefore must be emitted at a radius not larger than the deceleration 
radius of the blast wave. This radius is defined by\footnote{This definition 
assumes that half of $E$ is dissipated by the reverse shock in the fireball.}
$m=E/2c^2\Gamma^2$ where $E$ is fireball energy (left over after it emits 
the prompt GRB) and $m$ is the swept-up ambient mass. For a medium with 
density profile $n(R)\propto R^{-k}$, the mass within a radius $R$ is 
$m(R)=[4\pi/(3-k)]R^3nm_pc^2$, which gives
\be
\label{Rdec}
   \Rdec=\left[\frac{(3-k)E}{8\pi nm_pc^2\Gamma^2}\right]^{1/3}
   =1.4\times 10^{17}\,(3-k)^{1/3}E_{54}^{1/3}\ndec^{-1/3}\Gamma_2^{-2/3}
          \;{\rm cm},
\ee
\be
 \frac{\tIC(\nu)}{\texp}=6\times 10^{-3}\,\Gamma_2^{10/3}\nu_{15}^{-1/2}
   (1+z)^{-1/2} (3-k)^{1/3}E_{54}^{1/3}\epB^{1/4}\ndec^{-1/12}
   E_{\gamma,54}^{-1}\,\Delta_{12}\left(\frac{R}{\Rdec}\right)^{1-k/4},
\ee
where $\ndec=n(\Rdec)$.
Substituting here the observed parameters of GRB~990123, $z=1.6$,
$E_\gamma\approx 2\times 10^{54}$~erg, $\Delta\approx 10^{12}$~cm, one finds
that the electrons emitting in the optical band are slowly cooling if 
\be
\label{sc99}
   \Gamma>550\,(3-k)^{-1/10}\epB^{-3/40}\ndec^{1/40}
     \left(\frac{E}{E_\gamma}\right)^{-1/10}
    \left(\frac{R}{\Rdec}\right)^{-3/10(1-k/4)}
    \left(\frac{E_\gamma}{2\times 10^{54}}\right)^{1/5}
    \Delta_{12}^{-3/10}.
\ee
Since the flash radius cannot exceed $\Rdec$, we conclude that the 
slow-cooling condition can be satisfied in GRB~990123 if 
the Lorentz factor of the emitting region exceeds $500\epB^{-3/40}$.

An additional relation between $\Gamma$ and $R$ is given by the known 
arrival time of the flash, $t=(1+z)R/2\Gamma^2c\approx 50$~s. Combined 
with $R\leq\Rdec$ and condition~(\ref{sc99}), 
this gives a strong constraint on the deceleration radius 
$\Rdec>3.5\times 10^{17}(3-k)^{-1/5}\epB^{-3/20}
n^{1/20}(E/E_\gamma)^{-1/5}$~cm and ambient density,
\be
   \ndec<10^{-2}(3-k)^{3/2}\epB^{1/2}
      \left(\frac{E}{E_\gamma}\right)^{3/2}\; {\rm cm}^{-3}.
\ee

\subsection{Reverse-shock model}

The reverse shock can accelerate electrons with 
a power-law distribution and a mean Lorentz factor
\be
\label{gm}
   \bar{\gamma}_e=\frac{m_p}{m_e}\,\frac{\epe}{2}
   \left(\frac{\Gej}{\Gamma}+\frac{\Gamma}{\Gej}-2\right).
\ee
Here $\epe$ is the fraction of postshock energy density that is 
carried by the accelerated electrons, and $\Gej>\Gamma$ is the Lorentz 
factor of the preshock fireball. Since $\bar{\gamma}_e$ is comparable with 
$\gamma_e(\nu)$ for $\nu\sim 10^{15}$~Hz, the reverse shock is expected to 
be an efficient producer of optical radiation (M\'esz\'aros \& Rees 1993; 
Sari \& Piran 1999). 

The reverse-shock emission peaks at $R\approx\Rdec$ and then gradually decays
if the accelerated electrons cool down slowly, on the expansion timescale. 
Besides the slow-cooling conditions (eqs.~\ref{bound1} and \ref{sc99}), 
we note two other requirements:

\medskip

\noindent
1. --- The bulk of MeV photons can overlap with the reverse-shock emission 
as observed only if they are produced inside the fireball.
This agrees with the idea of internal dissipation as a source
of prompt $\gamma$-rays (e.g. Rees \& M\'esz\'aros 1994). 
The short-timescale variations in the prompt $\gamma$-rays indicate that 
they are produced at a radius $R_\gamma\ll\Rdec$. Then the MeV front 
gets strongly collimated by the time it reaches $\Rdec$
and propagates with velocity $c$ in the fireball frame.

\medskip
                                                                                
\noindent
2. --- The reverse shock can reach an emission peak before the 
MeV front fully overtakes it only if the shock is relativistic  ---
a non-relativistic shock would cross the fireball on a longer timescale 
and its emission would lag behind the $\gamma$-rays. 
This implies that $\Gej$ in GRB~990123 is even higher than required by 
the slow-cooling condition: $\Gej>2\Gamma>10^3\epB^{-3/40}$.

                                                                                
\section{GeV-TeV flash}

Inverse Compton scattering of MeV radiation in the flash region produces 
ultra-high-energy photons.
The prompt GRB spectrum usually peaks at $\ep_p=0.1-1$~MeV, which translates 
to $\ep_p/\Gamma\sim{\rm keV}$ in the fluid frame. This peak can be 
upscattered efficiently by electrons with 
\be
  \gamma_e<\gamma_e^*=\Gamma\,\frac{m_ec^2}{\ep_p}\sim \Gamma,
\ee
above which the Compton cross section is reduced by the 
Klein-Nishina correction.
The energy of upscattered photons $\epIC\sim \gamma_e^2\ep_p$ can 
extend to $\epIC^*\sim \Gamma^2\ep_p$ which is in the GeV-TeV range.

The upscattered photons will avoid $\gamma$-$\gamma$ absorption and escape 
the source if its optical depth $\taugg(\epIC)<1$. The optical depth seen 
by photons $\epIC=\epIC^*$ is 
\be
\label{tgg}
  \taugg(\epIC^*)\sim 0.1\,\frac{w_\gamma}{\ep_p}\,\sT R\approx 0.1 
                          \frac{\texp}{\tIC(\gamma_e^*)}.
\ee
For $\epIC<\epIC^*$, $\taugg$ is reduced as $(\epIC/\epIC^*)^\beta$ where 
$\beta$ is the slope of the prompt radiation spectrum at $\ep>\ep_p$.
Since $\gamma_e^*$ is not much different 
from $\gamma_e(\nu)$ for optical $\nu$, one concludes that a slow-cooling 
optical flash, $\tIC>\texp$, is also $\gamma-\gamma$ transparent. 

The emerging luminosity of ultra-high-energy (UHE) photons is much 
higher than the synchrotron luminosity if Compton losses dominate over 
synchrotron losses, i.e., if $w_\gamma\gg w_B$. 
The ratio of the two luminosities is given by
\be
  \frac{L_{\rm UHE}}{L_s}=\frac{w_\gamma}{w_B}
  \sim \frac{1}{\epB}\,
  \left(\frac{E_\gamma}{E}\right)\left(\frac{R}{\Rdec}\right)^{-2}.
\ee
(Here we assumed $\Rdec\sim 2\Gamma^2\Delta$ which is valid if the 
reverse shock is at least mildly relativistic). $L_{\rm UHE}$ is emitted 
as long as the flash overlaps with the MeV radiation front.
It ends when the blast wave begins to decelerate and the MeV front fully 
overtakes it.  

GRBs with Lorentz factors smaller than $10^3$ will have fast-cooling 
flashes, $\tIC<\texp$. Then the optical emission is suppressed. 
The fast Compton cooling is a possible reason of the non-detections 
of optical flashes in long GRBs observed by ROTSE.

In the fast-cooling case, $\tIC(\gamma_e^*)\ll \texp$,
the upscattered $\gamma$-rays may 
not avoid the $\gamma$-$\gamma$ absorption (see eq.~\ref{tgg}). Then an 
$e^\pm$ cascade develops from $\gamma_e^*$ to a lower $\gamma_{e,1}$ such 
that $\taugg(\ep_{\rm IC,1}=\gamma_{e,1}^2\ep_p)\sim 1$. The pairs 
resulting from this cascade cool to even lower $\gamma_{e,c}$ such that 
$\tIC(\gamma_{e,c})=\texp$. Most of the UHE luminosity should then 
emerge at energies $\sim\ep_{\rm IC,1}$.
The slope of the IC spectrum is $-1/2$ between 
$\ep_{\rm IC,c}\sim\gamma_{e,c}^2\ep_p$ and $\ep_{\rm IC,1}$
(the fast-cooling inverse-Compton spectrum). The slope below $\ep_{\rm IC,c}$ 
should approximately equal the slope of the prompt GRB spectrum at 
$\ep<\ep_p$.

If the flash emits a fraction $\epsilon_{\rm flash}$ of the fireball 
energy $E$, the ratio of the flash energy to the prompt GRB energy is 
$E_{\rm flash}/E_\gamma=\epsilon_{\rm flash}(E/E_\gamma)$.
This ratio can exceed unity if the radiative efficiency of the flash
exceeds that of the prompt GRB. 

The upscattered $\gamma$-rays have a collimation angle $\theta\sim 1/\Gamma$ 
and therefore lag behind the unscattered prompt radiation [which has a 
smaller collimation angle $\theta=(R_\gamma/\Rdec)\theta_\gamma$ where 
$R_\gamma$ is the radius of prompt emission and $\theta_\gamma$ is its 
initial collimation angle]. The resulting delay of high-energy $\gamma$-rays, 
$\delta t\sim (1+z)\Rdec/\Gamma^2$, is comparable with the duration of the 
prompt burst $t_b$, and the duration of the high-energy flash 
$t_{\rm UHE}\sim t_b+\delta t$ is a few times longer than $t_b$. 
The angular dispersion of high-energy 
photons also mixes up their arrival times and washes out short-timescale 
variability. This is a signature of upscattering at a large radius 
$\Rdec>R_\gamma$, which contrasts with the variable prompt radiation.

\subsection{GRB~941017}

A high-energy flash was observed in GRB~941017 by Compton Observatory
(Gonz\'alez et al. 2003). It lasted about 200~s, which is 2.5 times
longer than the prompt GRB, and had a hard spectral slope $\alpha\approx 0$
at $10-100$~MeV. Possible inverse Compton models were examined by 
Granot \& Guetta (2003) and found to be inconsistent with the data. 
The best proposed candidate was a reverse-shock emission (synchrotron 
self-Compton) which had the correct timing but still had a problem in 
explaining the spectral slope. Stern \& Poutanen (2004) considered 
prompt GeV emission from continuously heated electrons in the fireball
and Dermer \& Atoyan (2004) proposed a model involving acceleration 
of hadrons to ultra-high energies.

The data is consistent with the high-energy flash mechanism described 
above. All three expected features are observed: (1) the flash 
lasted a few times longer than the prompt MeV burst, (2) it did not show 
significant variability in the studied time bins, and (3) it 
peaked above the observed range $\ep<200$~MeV and the observed spectrum 
had approximately the same slope as the low-energy part of the prompt 
spectrum, consistent with the upscattering of the prompt 0.1~MeV photons.
The energy of the high-energy flash exceeds $E_\gamma$ by at least
a factor of 3, which points to a relatively low radiative efficiency 
during the prompt burst and a high radiative efficiency during the 
deceleration flash.

The data is consistent with an upscattering electron population 
that peaks at the cooling Lorentz factor $\gamma_{e,c}\sim 10$.
The electrons can be injected with higher $\gamma_e\sim 10^2-10^3$,
then undergo $e^\pm$ cascade to $\gamma_{e,1}\sim 10-100$, and
cool down to $\gamma_{e,c}\sim 10$. The peak of the high-energy 
spectrum is weakly constrained by the data, however, it is probably 
not far from 1~GeV --- otherwise the energy of the upscattered 
component would be very high.


\section{Conclusions}

When the optical flash overlaps with the prompt MeV front, like it 
does in GRB~990123, its main cooling mechanism is inverse 
Compton scattering of the MeV photons rather than synchrotron 
(or synchrotron self-Compton) emission. This strong cooling tends to 
terminate the flash.
Only if the fireball has a high $\Gamma\simgt 10^3$ is Compton cooling 
slow compared with the fireball expansion and consistent with the observed 
tail of the flash in GRB~990123. This condition can be translated to 
an upper bound on the ambient density $n\simlt 0.1\epB^{1/2}$~cm$^{-3}$
(eq.~11). If the flash was produced by the reverse shock in the fireball, 
the data requires the shock to be relativistic and the prompt MeV burst 
to originate inside the fireball.

In GRBs with typical Lorentz factors $\Gamma<10^3$, the flash is fast 
cooling as long as it overlaps with the MeV radiation.
The accelerated electrons quickly emit their energy by 
upscattering the MeV photons and produce 
a bright GeV flash, likely with the development of an $e^\pm$ cascade. 
The low-energy slope of the upscattered spectrum is 
the same as the low-energy slope of the prompt GRB, at 
higher energies it changes to $-1/2$ and is cut off by 
$\gamma-\gamma$ absorption in the source.
The temporal behavior of the high-energy flash differs 
from the prompt GRB: the short-timescale variability is washed out and the 
arrival time is extended by a factor of a few.
These features are observed in the high-energy component of GRB~941017.

One expects a clear spectral separation of the upscattered component
from the prompt $0.1-10$~MeV radiation. It is likely to peak well above 
1~GeV in most cases.
GRB~941017 appears to be a rare case where the peak is comparable with 
1~GeV, which makes the upscattered component well visible at $10-100$~MeV.
This rare case can be explained by a relatively low $\Gamma=100-200$ 
which places the cooling Lorentz factor at $\gamma_{e,c}\sim 10$.
The special character of this burst is confirmed by the fact that a similar 
component was not found in 25 other bursts studied by Gonz\'alez et al. 
(2003). GLAST should be able to observe the typical upscattered flashes 
at energies up to 100~GeV.

The overlapping of the decelerating fireball with MeV photons may not take 
place in all GRBs. The velocity of hot gas in the fixed frame is 
$c(1-1/2\Gamma^2)$, and the MeV front completely 
overtakes it at a radius $R_\Delta=2\Gamma^2\Delta$. 
The overlapping does not occur if 
$R_\Delta<\Rdec$, which requires the observed duration of the MeV front
\be
\label{dur}
  t_b<10\left(\frac{\Gamma}{300}\right)^{-2}R_{\rm dec,17}\;
  \frac{(1+z)}{2}\;{\rm s}.
\ee
The class of short GRBs with durations $t_b\sim 0.1$~s
can satisfy this condition and avoid the Compton cooling by 
MeV photons; the condition can also be met by long bursts with modest
$\Gamma$. Then an optical flash can be produced without a significant
GeV-TeV counterpart.\footnote{Condition~(\ref{dur}) implies
a non-relativistic reverse shock, and the prompt
photons leave the fireball before the shock crosses it.}

ROTSE observations of 3 short bursts impose upper limits on their optical 
flashes (Kehoe et al. 2001). The most stringent upper limit was obtained 
for GRB~980527 
($t_b=0.09$~s), where optical energy emitted at 15 seconds after the burst 
did not exceed $\sim 10^{-5}$ of the GRB energy. It can, however, be that 
the deceleration time for this burst was short, $\Rdec/\Gamma^2\ll 15$~s, 
and the flash was not seen because it was observed too late.

\acknowledgments
I am grateful to Chris Thompson for discussions and the referee 
for comments. This work was supported by NASA grant NAG5-13382.



\begin{references}

\reference{}
Akerlof, C. et al. 1999, Nature, 398, 400

\reference{}
Akerlof, C. et al. 2000, ApJ, 532, L25

\reference{}
Dermer, C. D., \& Atoyan, A. 2004, A\& A, 418, L5

\reference{}
Fox, D. W. 2002, GCN 1564 (http://gcn.gsfc.nasa.gov/gcn/gcn3/1564.gcn3) 

\reference{}
Fox, D. W. et al. 2003, ApJ, 586, L5 

\reference{}
Gonz\'alez, M. M., Dingus, B. L., Kaneko, Y., Preece, R. D.,
Dermer, C. D.,  \& Briggs, M. S. 2003, Nature, 424, 749 

\reference{}
Granot, J., \& Guetta, D. 2003, ApJ, 598, L11

\reference{}
Kehoe, R. et al. 2001, ApJ, 554, L159

\reference{}
Kobayashi, S. 2000, ApJ, 545, 807

\reference{}
M\'esz\'aros, P., \& Rees, M. J. 1993, ApJ, 418, L59

\reference{}
Nakar, E., \& Piran, T. 2004, MNRAS, 353, 647

\reference{}
Rees, M. J., M\'esz\'aros, P. 1994, ApJ, 430, L93

\reference{}
Sari, R., \& Piran, T. 1999, ApJ, 517, L109

\reference{}
Stern, B., \& Poutanen, J. 2004, MNRAS, 352, L35
\end{references}
\end{document}